\documentclass[showpacs,amssymb,superscriptaddress,notitlepage,nofootinbib,prd,longbibliography]{revtex4-2}
\usepackage{graphicx}
\usepackage{amsmath}
\usepackage{amsfonts}
\usepackage{bm}
\usepackage[colorlinks=true]{hyperref}
\usepackage[all]{hypcap}
\usepackage[utf8]{inputenc}
\usepackage{url}
\usepackage{epigraph}

\usepackage{float}

\usepackage{xcolor}
\usepackage{multirow}
\usepackage{ebgaramond}
 \usepackage{etoolbox}

\hypersetup{
    colorlinks,
    citecolor=[rgb]{0.16,0.53,0.80},
    linkcolor=[rgb]{0.16,0.03,0.50},
}

\input colordvi
\usepackage{tikz}
\usepackage{graphicx,color}

\def\be{\begin{equation}}
\def\ee{\end{equation}}
\def\bea{\begin{eqnarray}}
\def\eea{\end{eqnarray}}
\def\ptl{\partial}

\begin{document}

\renewcommand\textflush{flushright}

    \makeatletter
    \newlength\epitextskip
    \pretocmd{\@epitext}{\em}{}{}
    \apptocmd{\@epitext}{\em}{}{}
    \patchcmd{\epigraph}{\@epitext{#1}\\}{\@epitext{#1}\\[\epitextskip]}{}{}
    \makeatother

    \setlength\epigraphrule{0pt}
    \setlength\epitextskip{1ex}
    \setlength\epigraphwidth{.4\textwidth}

\thispagestyle{empty}

\epigraph{\it\small Wo suchen wir das Nichts? Wie finden wir das
Nichts? M{\"u}ssen wir, um etwas zu finden, nicht {\"u}berhaupt
schon wissen, da{\ss} es da ist?} {\it\small Martin Heidegger,
``Was ist Metaphysik''} \epigraph{\it\small Everything not
forbidden is compulsory.} {\it\small Murray Gell-Mann}

\centerline{\Large{ \bf \AE ther as an Inevitable Consequence of
Quantum Gravity }}
\medskip
\centerline{S.L. Cherkas $^a$, V.L. Kalashnikov$^b$}
\medskip
\centerline{$^a$Institute for Nuclear Problems, Belarus State
University} \centerline{Minsk 220006, Belarus} \centerline{Email:
cherkas@inp.bsu.by}
\medskip
\centerline{$^b$ Department of Physics,}\centerline{  Norwegian
University of Science and
Technology,}\centerline{H{\o}gskoleringen 5, Realfagbygget, 7491,
Trondheim, Norway} \centerline{Email:
vladimir.kalashnikov@ntnu.no}
\medskip
\centerline{\small Submitted: November 29, 2022}

\smallskip

{\fontsize{10}{11}\selectfont The fact that quantum gravity does
not admit an invariant vacuum state has far-reaching consequences
for all physics. It points out that space could not be empty, and
we return to the notion of an \ae ther. Such a concept requires a
preferred reference frame for describing universe expansion and
black holes. Here,  we intend to find a reference system or class
of metrics that could be attributed to ``\ae ther''. We discuss a
vacuum and quantum gravity from three essential viewpoints:
universe expansion, black hole existence, and quantum
decoherence.}

\section{Introduction}

\begin{figure}[H]
\hspace{3 cm}
\includegraphics[width=12cm]{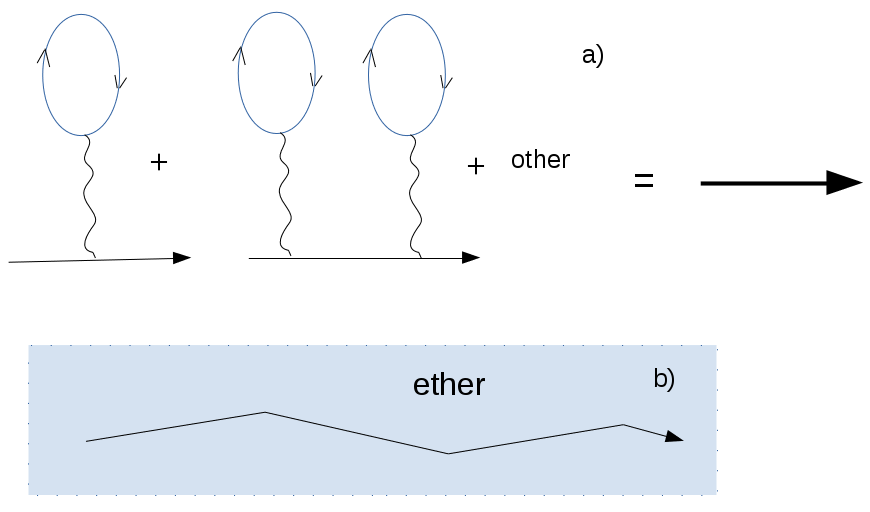}
\vspace{0 cm} \caption{Illustration of vacuum influence on the
particle propagation in (\textbf{a}) QFT, where the vacuum loops
renormalize the mass and charge of the particle but do not prevent
their free motion, and in \linebreak (\textbf{b}) QG, where the
\ae ther fills space due to the absence of an invariant vacuum
state. }
\label{vv}
\end{figure}

From the earliest times, people comprehended an emptiness as
``Nothing'', which consists of absolutely nothing, no matter, no
light, nothing. Others are convinced that ``nothing'' is
unthinkable and a space-time should always contain ``something,''
i.e., to be ``\ae ther''  \cite{Schaffner72}. From straightforward
point of view, the \ae ther represents some stationary ``medium''
mimicking some matter and needs a preferred reference frame in
which it is at rest ``in tote''. After the development of the
quantum field theory (QFT), it was found that a vacuum actually
contains a number of virtual particle--antiparticle pairs
appearing and disappearing during the time of $\Delta
t\propto\frac{1}{m}$, where $m$ is a particle mass. That leads to
the experimentally observable effects such as anomalous electron
magnetic moment, the Lamb shift of atomic levels
\cite{Landau1982}, the Casimir effect
\cite{klimchitskaya2006experiment}, etc.  However, although a
vacuum is not empty, a ``soup'' of the virtual
particle--antiparticle pairs is not \ae ther because it does not
prevent the test particles from moving freely due to the Lorentz
invariance (LI) of a QFT vacuum, as it is illustrated in Figure
\ref{vv}. That implies  rigid limits on a local LI violation, and
the existence of a preferred reference frame in the framework of
QFT \cite{mattingly2005modern}. However, considering gravity seems
to insist on the \ae ther existence and the preferred reference
frame due to an absence of a vacuum state invariant relative
general transformation of coordinates. That demands reconsidering
an idea of \ae ther \cite{dirac1951there}. A possibility of the LI
violation was also considered within string theory and loop
quantum gravity (see \cite{collins2006lorentz} and references
herein), the Einstein-\AE ther \cite{aether}, and Horava--Lifshitz
\cite{Hor} theories, and others (see \cite{h1,h2} for
phenomenological implications). It could also mention the CPT
invariance violation \cite{anca}, which manifests itself both
under Minkowski's space-time \cite{Cherkas2002,Kostel2011} and in
the presence of gravity \cite{Kostel2018,h2}.

{Another argument for a preferred reference frame is the vacuum
energy problem. If the zero-point energy is real, we need to
explain why this energy does not influence a universe's expansion.
One of the solutions is to modify the gravity theory. That may
violate the invariance relative to the general transformation of
coordinates. For example, the Five Vectors Theory (FVT) of gravity
demonstrates such a violation, including a LI violation
\cite{Vesti}.}

\section{Vacuum State and QG}

The notion of a vacuum state originates from the ground state of a
quantum oscillator. In QFT, the free fields are decomposed into a
set of independent field oscillators by Fourier decomposition.
Exited states of the oscillators are treated as particles, i.e.,
matter (both massive and massless). Introducing the interaction
term leads to the renormalization of a particle mass and charge,
but a one-particle state remains a one-particle state.
Consequently, a one-particle wave packet moves freely with the
constant envelope velocity, i.e., with no in vacuo dispersion
\cite{amelino1998tests}. That implies that even in the presence of
perturbative interaction, one could still introduce a LI vacuum
state in QFT.

Quantization of GR is too complicated to discuss a vacuum state.
Nevertheless, let us consider a toy QG model regarding this issue.
In this model only a spatially nonuniform scale factor represents
gravity $a(\eta,\bm r)$. It is certainly not a self-consistent
approach within the GR frameworks \cite{toy}. Nevertheless, there
exists a (1+1)-- dimensional  toy model  \cite{toy} including a
scalar fields $\bm
\phi(\tau,\sigma)=\{\phi_1(\tau,\sigma),\phi_2(\tau,\sigma)
\dots\}$ and a scale factor $a(\tau,\sigma)$ described by the
action\vspace{-6pt}
\begin{equation}   S=\int L\,d\tau=\frac{1}{2}\int \left(-a^{\prime
2}+(\partial_\sigma a )^2+a^2\left(\bm \phi^{ \prime
2}-(\partial_\sigma{\bm \phi})^2\right)\right)d \sigma d\tau,
\label{mod}
\end{equation}
where $\tau$ is a time variable, $\sigma$ is a spatial variable,
and prime denotes differentiation with respect to $\tau$. Here,
like GR, the scalar fields evolve on the curved background
$a(\tau,\sigma)$, which is, in turn, determined by the fields. The
 equations of motion is written as
\begin{equation}
 \bm \phi^{\prime\prime}-\partial^2_{\sigma}\bm
\phi+2\alpha^\prime\bm \phi^\prime
-2\partial_{\sigma}\alpha\partial_{\sigma}\bm
\phi=0,~~~~~~~~~~~~~~~~ ~~\label{eqns10}
\end{equation}
\begin{equation}
\alpha^{\prime\prime}-\partial^2_{\sigma}\alpha+\alpha^{\prime
2}-(\partial_\sigma\alpha)^2+\bm \phi^{\prime
2}-(\partial_\sigma\bm \phi)^2=0.
\label{eqns1}
\end{equation}

The relevant Hamiltonian and momentum constraints, written in
terms of momentums $\bm \pi(\tau,\sigma)\equiv\frac{\delta
L}{\delta {\bm \phi^\prime(\tau,\sigma) }}=a^2\bm \phi^\prime$,
$p_a(\tau,\sigma)\equiv-\frac{\delta L}{\delta a
^\prime(\tau,\sigma)}=a^\prime$ is\vspace{-6pt}

\begin{equation}
\mathcal{H}=\frac{1}{2}\left(-p_a^{2}+\frac{\bm \pi^{
2}}{a^2}+a^2(\partial_\sigma{\bm \phi})^2-(\partial_\sigma
a)^2\right)=0,\label{sv0}
\end{equation}
\begin{equation}
\mathcal{P}=-p_a\partial_\sigma a+ \bm \pi\partial_\sigma\bm
\phi=0,
\label{sv}
\end{equation}
 and obey the constraint evolution similar to GR \cite{toy}:\vspace{-6pt}
 \begin{equation}
\partial_\tau{\mathcal H}=\partial_\sigma{\mathcal P},
\end{equation}
\begin{equation}
\partial_\tau {\mathcal P}=\partial_\sigma {\mathcal H}.
\label{prop} \end{equation}

\subsection{  {Quasi-Heisenberg Quantization and a Region of Small Scale Factor: Absence of Vacuum State}}
\label{sub21}

It is believed that our universe originates from a singularity in
which a scale factor equals zero. Let us consider a region of
small scale-factors first. In this region, it is convenient to use
the quasi-Heisenberg picture \cite{cherkas2005quantum}, in which
the setting of the initial conditions for operators at the initial
moment allows quantization of the equations of motion. In the
vicinity of small-scale factors, kinetic energy terms dominate
over potential ones  \cite{toy,cherkas2005quantum} so that the
equations of motion (\ref{eqns10}) {and} 
 (\ref{eqns1}) reduce to\vspace{-6pt}
\begin{equation}
 \hat{\bm \phi} ^{\prime\prime}+2 \hat{\alpha}^\prime\hat{\bm \phi} ^\prime
\approx 0, ~~\label{ev1}
\end{equation}
\begin{equation}
\hat{\alpha}^{\prime\prime}+\hat{\alpha}^{\prime 2}+\hat{\bm
\phi}^{\prime 2}\approx 0.
\label{ev2}
\end{equation}

The solutions of Equations (\ref{ev1}) {and} (\ref{ev2}) for two
scalar fields $\phi_1(\tau,\sigma)$, $\phi_2(\tau,\sigma)$ under
initial conditions, discussed in Appendix \ref{quasi}, are written
as\vspace{-6pt}
\begin{eqnarray*}
\hat { \phi_1}(\tau,\sigma)= -\frac{i}{\pi_1}\int_{-\infty}^\infty
\theta(\sigma-\sigma^\prime)S\left(k(\sigma^\prime)\ptl_{\sigma^\prime}\,
\frac{\delta}{\delta
k(\sigma^\prime)}\right)d\sigma^\prime+~~~~~~~~~~~~~~\nonumber\\\frac{
 \pi_1}{2\sqrt{\pi_1^2+k^2(\sigma)}}\ln\left(1+2e^{-2 \alpha_0
}\sqrt{\pi_1^2+k^2(\sigma)}\,\tau\right),
\\
\hat {\phi_2}(\tau,\sigma)=i\frac{\delta}{\delta k(\sigma)}+\frac{
k(\sigma)}{2\sqrt{\pi_1^2+k^2(\sigma)}}\ln\left(1+2e^{-2 \alpha_0
}\sqrt{\pi_1^2+k^2(\sigma)}\,\tau\right),\\
\hat \alpha(\tau,\sigma)=\alpha_0+\frac{1}{2}\ln\left(1+2\,e^{-2
\alpha_0 }\sqrt{\pi_1^2+k^2(\sigma)}\,\tau\right),
\end{eqnarray*}
where the notations are given in the Appendix \ref{quasi}.

As one can see, the scalar fields and the logarithm of the scale
factor have monotonic behavior with time. It means that there are
no oscillators in the vicinity of small-scale factors and no
possibility of defining a vacuum state.   In this situation, a
quantum state  is described by the momentum wave packet
$C[k(\sigma)]$  as it is discussed in the Appendix \ref{quasi}.

The difference in behavior in the vicinity of small-scale factors
and at the epoch of the quantum oscillators occurrence was known
long ago from analysis of the Wheeler--DeWitt equation solutions
for the Gowdy model \cite{mizng}. Therefore, we simply illustrate
this fact in terms of asymptotic solutions of operator equations.

 \subsection{String-Like Quantization within the Intermediate Region}
 \label{sub22}

 From the previous subsection
 one can see that the operator equations of motion are not the oscillator equations in the vicinity of \emph{a}$\sim$0,
 which does not allow defining a vacuum state. A question arises: Could we define a vacuum state when the fields begin to
 oscillate, and quantum oscillators arise? In this region, the fields obey
 nonlinear wave \linebreak Equations (\ref{eqns10}) and (\ref{eqns1}),
 which could not be solved analytically. That complicates using
 the quasi-Heisenberg picture, and to obtain some analytical results, we will use bosonic string quantization \cite{green,kaku}.    The action (\ref{mod}) could be rewritten in the reparametrization invariant form of a string on the curved\vspace{-6pt}
background \cite{toy}
\begin{equation}
\label{action}
S=\frac{1}{2}\int
d^2\xi\sqrt{-g}\,g^{\alpha\beta}(\xi)\partial_\alpha
X^A\partial_\beta X^B G_{A\,B} (X(\xi)),
\end{equation}
where $\xi=\{\tau,\sigma\}$, $X^A=\{a,\phi_1,\phi_2,\dots\}$, and
the metric tensors $g_{\mu\nu}$, $G_{A\, B}(X)$ are in the
\linebreak form of\vspace{-6pt}
\[g= \left(
\begin{array}{cc}
-N^2+N_1^2&  N_1
\\
N_1 &1
\end{array}
\right),~~~~~~~~~~~~~~~G= \left(
\begin{array}{cccc}
1  &0&0&\dots
\\
0 &-a^2&0&\dots
\\
0 &0&-a^2&\dots
\\
\dots &\dots&\dots&\dots
\end{array}
\right).
 \]

 The particular gauge for the lapse  $N=1$ and shift
$N_1=0$ functions results in (\ref{mod}). The metric tensor
$g_{\alpha\beta}(\xi)$ describes an intrinsic geometry of a
(1+1)-dimensional manifold, i.e., a (1+1)-dimensional space-time,
and it is an analog of the four-dimensional metric of general
relativity. $G_{A\, B}(X(\xi))$ represents a geometry of the
external space unifying scale factor and scalar fields and has no
direct physical meaning here. The system (\ref{action}) manifests
an invariance relative to the reparametrization of the variables
$\tau, \sigma$, which is analog of the general coordinate
transformation in GR.  The transformations of coordinates $\tilde
\tau=\tilde \tau(\tau,\sigma)$,
 $\tilde \sigma=\tilde \sigma(\tau,\sigma)$ imply transition to another reference frame for an observer who ``lives on a~string''.

 For obtaining a vacuum state, the key point is fixing the gauge by taking
$g_{\mu\nu}$ in the form of Minkowski's metric by setting $N=1$,
$N_1=0$, which simplifies the action (\ref{action}) to
 the form
\begin{equation}
S=\frac{1}{2}\int d\sigma d\tau G^{AB} \left(-\ptl_\tau
X_B\ptl_\tau X_A+\ptl_\sigma X_B\ptl_\sigma X_A\right).
\end{equation}

The momentum\vspace{-6pt}
\begin{equation}
P^A=\frac{\delta S}{\delta(\ptl_\tau X_A)}=-\ptl_\tau
X^A=-G^{AB}\ptl_\tau X_A
\end{equation}
 and
 the variable $X_A$ obey the canonical
commutation relations
\begin{equation}
[\hat P_A(\tau,\sigma),\hat X_B(\tau,\sigma^\prime)]=i
G_{AB}\delta(\sigma-\sigma^\prime).
\label{canon}
\end{equation}

As a zero-order approximation, one may take $G$ to be equal to
$\mathcal G$, where
\[\mathcal G= \left(
\begin{array}{ccc}
1  &0&\dots
\\
0 &-1&\dots
\\
\dots &\dots&\dots
\end{array}
\right).
 \]

 Then, it could be possible to develop the perturbation theory
on $G-\mathcal G$. In zero-order, $X_A$ satisfies  the wave
equation
\begin{equation}
\hat X_A^{\prime\prime}-\ptl^2_{\sigma}X_A=0,
\label{wave}
\end{equation}
and the commutation relations (\ref{canon}) can be realized using
creation and annihilation operators\vspace{-6pt}
\begin{equation}
\hat X_A=\sum_{k=-\infty}^\infty
\frac{1}{\sqrt{2k}}\left(\mbox{a}_{kA}e^{ik\sigma-i|k|\tau}+\mbox{a}_{kA}^+e^{-ik\sigma+i|k|\tau}\right),
\end{equation}
\begin{equation}
\hat P_A=\sum_{k=-\infty}^\infty
i\sqrt{\frac{k}{2}}\left(-\mbox{a}_{kA}e^{ik\sigma-i|k|\tau}+\mbox{a}_{kA}^+e^{-ik\sigma+i|k|\tau}\right),
\end{equation}
where $\mbox{a}_{kA}$, $\mbox{a}_{kB}^+$ obey
\begin{equation}
[\mbox{a}_{kA},\mbox{a}_{qB}^+]=-\mathcal G_{AB}\delta_{k,q}.
\label{comm1}
\end{equation}

Thus, only when the gauge is fixed by $N=1$, $N_1=0$, it is
possible to define a vacuum state by $\mbox{a}_{kA}|0>=0$. This
vacuum state is not gauge-invariant because the dynamic variable
$X_A$ satisfies the wave Equation (\ref{wave}) in only this gauge
(and in zero-order on $G-\mathcal G$). Moreover, one could see a
problem with the definition of the Fock space of quantum states.
Actually, Equation (\ref{comm1}) leads to
 $[\mbox{a}_{k0},\mbox{a}_{k0}^+]=-1$. That means that the state
 $\mbox{a}_{k0}^+|0>$ has a negative norm $<0|\mbox{a}_{k0}\mbox{a}_{k0}^+|0>=-1$. To avoid the negative norms, the string theory uses additional conditions on the physical Fock states $|>$:\vspace{-6pt}
\begin{equation}
\hat {\mathcal L}_f|>=0,
\end{equation}
 where $\hat {\mathcal L}_f=\int (\hat P_A(\tau,\sigma)+\ptl_\sigma
 \hat X_A(\tau,\sigma))^2\,f(\sigma)d\sigma$, and $f(\sigma)$ is an arbitrary function. Operators $\hat {\mathcal L}_f$ obey the Virasoro
 algebra.   It should be noted that the definition of the Virasoro operators includes the normal ordering \cite{green,kaku,Kiri}, but it is beyond the concept of our work. If one accepts the feasibility of using the normal ordering, then the vacuum energy problem does not exist at all.  However, we intend to refrain from discussing the status of excluding anomalies in the string theory here.

\subsection{Towards a Classical Background}
  {In Section \ref{sub21}, it is shown that there is no
vacuum state in the vicinity of a small scale-factor because of an
absence of field oscillators. In principle, the quasi-Heisenberg
picture could be used for the description of the subsequent
evolution, but it could be done only numerically because solving
the operator equations with the initial conditions is complicated.
Instead, we have used a string-like quantization described in
Section \ref{sub22}. That allows an analytical consideration of
the vacuum state, but it is only half of the problem because a
further investigation of the perturbation series on $G-\mathcal G$
is needed. Moreover, the trouble with the negative norm of the
states can be solved based on the Virasoro algebra by the
transition to the $D=26$ dimension in the string theory
\cite{green,kaku,Kiri}. The general conclusion for us is that the
vacuum state is not gauge-invariant and is defined in a single
gauge $N=1$, $N_1=0$. We could not make some other physical
predictions for this region. However, one could put forward a
hypothesis that in the presence of multiple scalar fields, a scale
factor acquires monotonic behavior in time and could be considered
classically finally. Such a situation is studied in the next
section and allows for obtaining a number of physical
predictions.}

\section{Vacuum Energy Problem as a Criterion for Finding the Preferred Reference Frame}

The more straightforward problem is to define the vacuum state on
a classical background space-time. Even in this case, the exact
vacuum state exists only for some particular space-time. In other
cases, the vacuum state has only an approximate meaning
\cite{anischenko2008functional}. The observer moving with
acceleration straightforwardly \cite{Unruh} or circularly
\cite{Akhmedov2008} in Minkowski's space-time will detect quanta
of the fields. That means that, although an observer could be in a
resting coordinate system, the quantum fields are not in a vacuum
state.

Nevertheless, a vacuum state could be defined, for example, in the
slowly expanding universe, where a solution to
 the vacuum energy problem could serve as a criterion for
 choosing a preferred reference frame.
The solution implies avoidance of the enormous zero-point energy
density of the quantum fields affecting the universe's expansion.
To do this, a class of conformally unimodular (CUM) metrics has
been introduced \cite{Vesti}:\vspace{-6pt}
\begin{equation}
  ds^2\equiv g_{\mu\nu} dx^\mu dx^\nu = a^2\left(1-\ptl_m
P^m\right)^2d\eta^2-\gamma_{ij} (dx^i+ N^i d\eta) (dx^j+
N^jd\eta),
\label{interv1}
\end{equation}
where $x^\mu=\{\eta,\bm x\}$, $\eta$ {is} 
 a conformal time,
$\gamma_{ij}$ is a spatial metric, $a =\gamma^{1/6}$ is a locally
defined scale factor, and $\gamma=\det\gamma_{ij}$.  The interval
(\ref {interv1}) is similar formally to the ADM one \cite{adm},
but the lapse function is taken in the form of $a(1-\ptl _ m P ^
m) $, where $P ^ m $ is a three-dimensional vector, and $\ptl_m$
is a conventional partial derivative.

Using the restricted class of the metrics (\ref{interv1}), the
theory \cite{Vesti} has been suggested in which the Hamiltonian
constraint is not necessarily zero but equals some constant. Such
a theory is known as the Five Vectors theory (FVT) of gravity
\cite{Vesti}, because the interval (\ref{interv1}) contains two
3-vectors {$\bm P$, $\bm N$} 
 and, moreover, spatial metric can be
decomposed into a set of three triads $\gamma_{ij}=e_{ia}e_{j a}$,
where index $a$ enumerates vectors of the triads $\bm e_{a}$.

This theory satisfies the strong equivalence principle (EP)
because no additional tensor fields appear.\footnote{{See} 
\cite{pit,KnoxStudies} for EP historical and philosophical
overview, and \cite{comp} for compatibility of EP with QFT.}
Nevertheless, in contrast to GR, where the lapse and shift are
arbitrary functions fixing the gauge, the restrictions $\ptl _
n(\ptl _ m N^m )=0$ and $\ptl _ n(\ptl _ m P^m )=0$ arise in FVT.
The Hamiltonian $\mathcal H$ and momentum $\mathcal P_i$
constraints in  the particular gauge $P^i=0$, $N^i=0$  obey the
constraint evolution equations \cite{Vesti}:\vspace{-6pt}
\begin{equation}
\ptl_\eta{\mathcal H}=\ptl_i\left(\tilde \gamma^{ij}\mathcal
P_j\right),\label{5}
\end{equation}
\begin{equation}
\ptl_\eta {\mathcal P_i}=\frac{1}{3}\ptl_i {\mathcal H},\label{6}
\end{equation}
where $\tilde \gamma_{ij}=\gamma_{ij}/a^2$ is a matrix with a unit
determinant. Equations (\ref{5}) and (\ref{6}) admit adding some
constant to ${\mathcal H}$ and, in the FVT frame, it is not
necessary that $\mathcal H=0$, but  $\mathcal H=const$ is also
allowed. That solves the problem of the main part of the
zero-point energy density.

Let us consider a spatially uniform, isotropic, and a flat
universe with the metric\vspace{-6pt}
\begin{equation}
ds^2=a(\eta)^2(d\eta^2-d\bm x^2),
\label{m1}
\end{equation}
which belongs to a class of (\ref{interv1}). Using the Pauli hard
cutoff of the 3-momentums \linebreak $k_{max}$
\cite{Pauli1971,visser} reduces the zero-point energy density
calculated in the metric (\ref{m1}) to
\begin{equation}
    \begin{split}
        \rho_v &= \frac{(N_{boson}-N_{ferm})}{4 \pi^2a^4}\int_0^{k_{max}}k^2\sqrt{k^2+a^2m^2} dk \approx  \\
&\frac{(N_{boson}-N_{ferm})}{16\pi^2}\biggl(\frac{k_{max}^4}{a^4}+\frac{
m^2 k_{max}^2}{a^2}
  + \frac{m^4}{8}\left[1+2\ln\left(\frac{m^2a^2}{4k_{max}^2}\right)\right]+\ldots \biggr),~
    \end{split}
    \label{mass0}
\end{equation}
where, for simplicity, bosons and fermions of equal masses are
considered.

 The main part of this energy density
$\sim\frac{k_{max}^4}{a^4}$ scales as radiation, and it has to
cause an extremely fast universe expansion in the frame of GR.
This result contradicts the observations \cite{dol}. In our
approach, a constant in the Hamiltonian constraint \cite{Vesti}
compensates this main part of zero-point energy and makes it
unobservable.\footnote{It should be noted that a mutual
cancellation of the bosonic $N_{boson}$ and fermionic $N_{boson}$
degrees of freedom removes all the vacuum energy but demands exact
supersymmetry, which was not observed to date \cite{super}.}

The remaining parts in (\ref{mass0}) are also huge but assuming
the sum rules for masses of bosons and fermions (the condensates
should be taken into account, as well) would provide a mutual
compensation for these terms \cite{visser,equation}. Of course,
all spectrum of the particles in nature, including unknown now,
should be taken into account. The empirical cutoff of momentums
$k_{max}$ is used in (\ref{mass0}), with the hope that some
fundamental basis will be found for that in the future (e.g., like
a noncommutative geometry \cite{n1,n2,n3}), and will provide the
UV completions of QG without a renormalization.

 Equation (\ref{interv1}) determines the preferred reference
 frame ensuring an \ae ther existence and an absence
of dipole anisotropy of the cosmic microwave background (CMB)
\cite{dodel}. Otherwise, the question arises: What is the physical
foundation of the frame where CMB is in a rest ``in tote,'' i.e.,
does not have a dipole component
\cite{universe7080311}?\vspace{-6pt}

\section{Cosmological Consequences of Residual Vacuum Energy}

\begin{figure}[H]
    \includegraphics[width=9.cm]{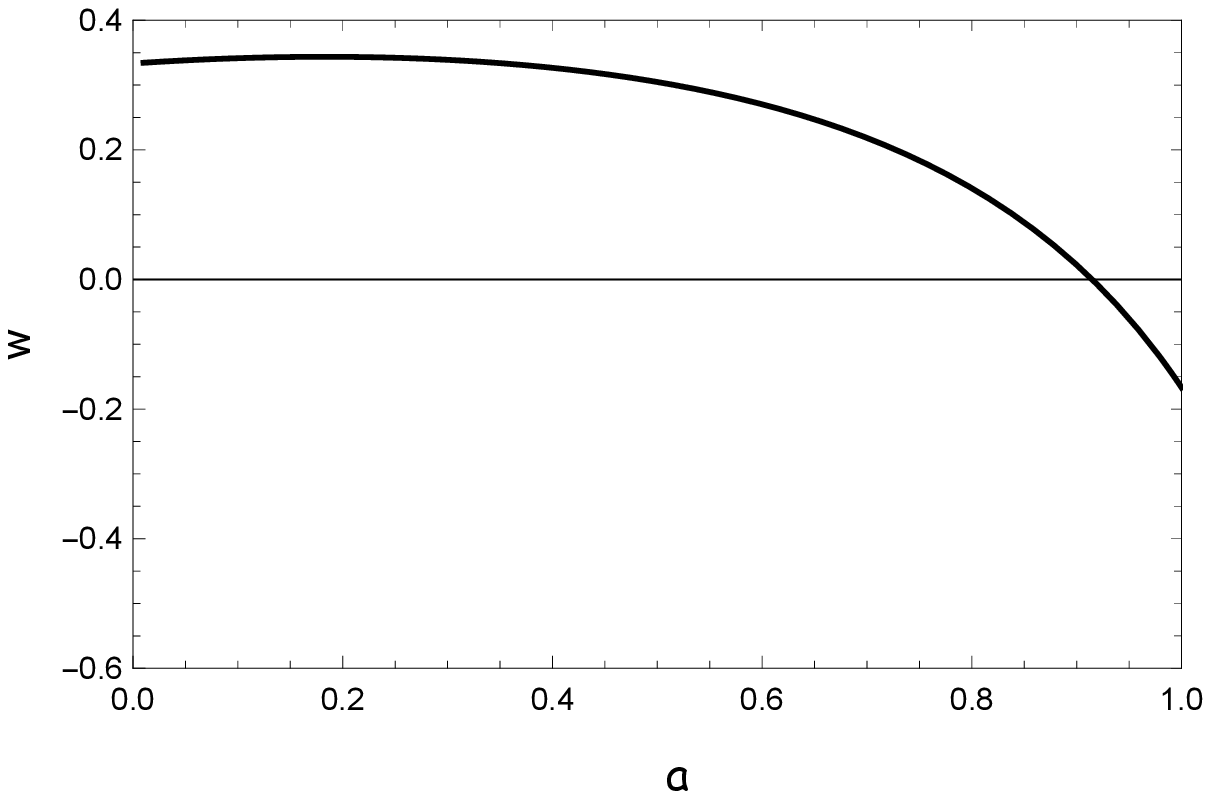}\\
\includegraphics[width=11.cm]{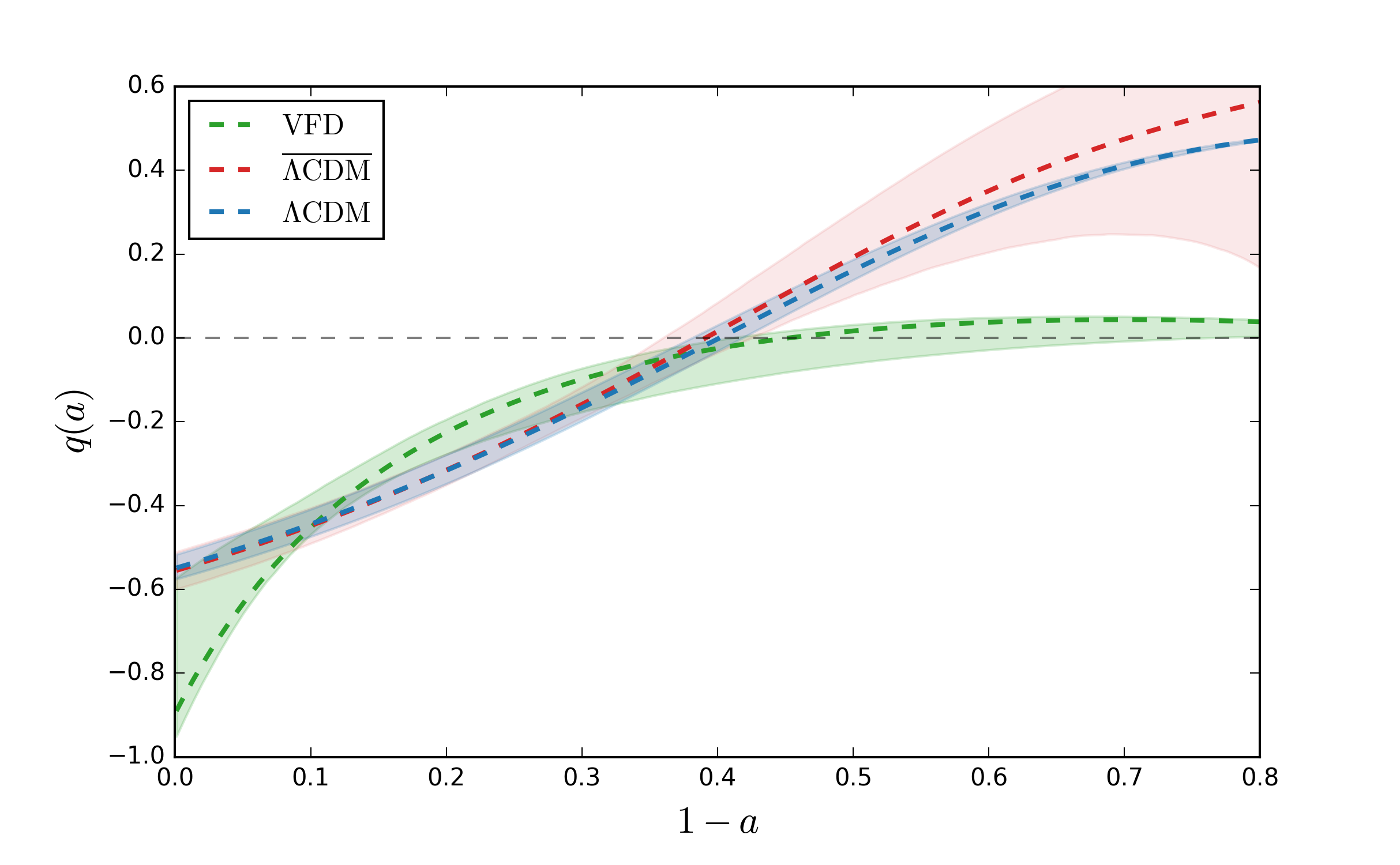}
    \caption{(\textbf{{Upper}
 panel}).
 {Equation}
  of the vacuum state in dependence
    on the universe scale factor $a$. (\textbf{{Lower panel}}) Deceleration parameter $q(a)$ and
     the corresponding
    dispersion channels for the VFD model (\ref{eqn:rhoandp}) and two versions of $\Lambda$CDM model. }
    \label{fig1}
\end{figure}

Other contributors to the vacuum energy density are the terms
depending on the derivatives of the universe expansion rate
\cite{Cherkas2007,Cherkas2008,equation}. Sum rules cannot remove
these terms, but they have the correct order of $\rho_{v}\sim
M_p^2 H^2$, where $H$ is the Hubble constant, and allow explaining
the accelerated expansion of the universe. These energy density
and pressure are
\cite{Cherkas2007,Cherkas2008,equation}:\vspace{-6pt}
\begin{align}
    \label{eqn:rhoandp}
\rho_v&=\frac{a^{\prime
2}}{2a^6}M_p^2S_0,~~~~~~~~p_v=\frac{M_p^2S_0}{a^6}\left(\frac{1}{2}a^{\prime
2}-\frac{1}{3}a^{\prime\prime}a\right),
\end{align}
where, $
    S_0 = \frac{k_{max}^2}{8 \pi^2 M_p^2}
$ is determined by the UV cut-off of the comoving momenta and the
reduced Planck mass $M_p=\sqrt{\frac{3}{4\pi G}}$ is implied. The
energy density and pressure of vacuum (\ref{eqn:rhoandp}) satisfy
a continuity equation\vspace{-6pt}
\begin{equation}
    \rho_v^\prime+3\frac{a^\prime}{a}(\rho_v+p_v)=0,
    \label{3}
\end{equation}
and, in the expanding universe,  are related to the equation of
state $p_{v}=w\,\rho_{v}$, as Figure \ref{fig1} (upper panel)
illustrates. Using this equation of vacuum state leads to the
cosmological Vacuum Fluctuations Domination (VFD) model
\cite{Haridasu,Cherkas2007,Cherkas2008}. According to VFD the
universe behavior at early times, when the scale factor was small,
is as freely rolling, i.e., without any deceleration or
acceleration, but it is accelerated at a late time. The
deceleration parameter
$q(a)=-\frac{a^{\prime\prime}a}{{a^\prime}^2}+1$  is shown in
{Figure}
 \ref{fig1} (lower panel) \cite{Haridasu}. The discovery of an
accelerated universe expansion was a big surprise \cite{perl}.
However, if the above view of a vacuum is true, a stage preceding
the acceleration should be Milne-like, i.e., linear in a cosmic
time. The Milne-like universes  have been much discussed again
recently
\cite{sul,klinkhamer2019instability,Wan2019,john2019r,Manfredi2020,lewisbarnes2021,chardin2021mondlike}.

\subsection{Nucleosynthesis in the Milne-like universe}

\begin{figure}[H]
\includegraphics[width=13.5cm]{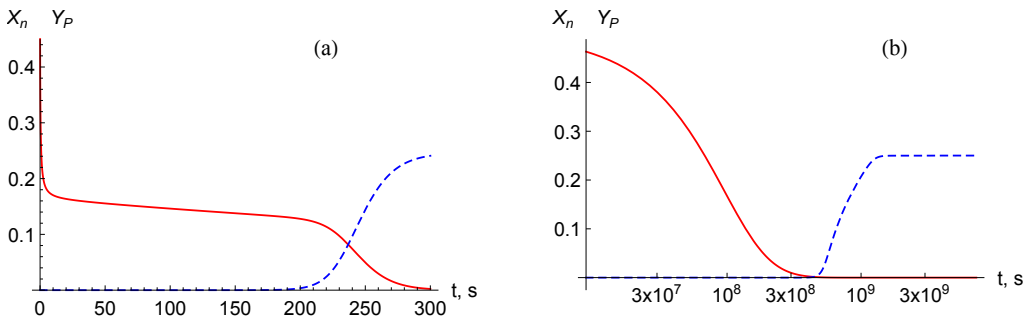}
\vspace{0.0 cm} \caption{Dependencies of relative abundances of
neutron and $^4\mbox{He}$ on cosmic time $dt=ad\eta$ are given by
red and  blue curves, respectively, (\textbf{a}) for standard
cosmological model, (\textbf{b}) for the VFD model.}
\label{nucl}
\end{figure}

Nucleosynthesis in a slowly expanding universe was considered
earlier \cite{Batra,Singh19,Lewis16}. Here, we present our
calculation for the VFD model, which has a Milne-like stage, as
shown in Figure \ref{fig1}, corresponding to the region where the
deceleration parameter $q$ is close to zero. The calculations have
been performed with the PRIMAT code (version 0.1.1)
\cite{primat,Pitrou2021} including 423 nuclear reactions. The
results of the calculations are presented in Table \ref{nn1} and
Figure \ref{nucl}.  For comparison, the results for the standard
cosmological model are also shown.

As expected, there is a very low rate of neutrons during a period
of helium formation in the VFD model (see Figure \ref{nucl}b).
That is because an equilibrium between protons and neutrons is
shifted towards a neutron decay during the slow universe
expansion.
 Nevertheless, a small amount
of neutrons during a long time can create a necessary amount of
helium if
 baryonic density $\Omega_b\approx 0.76$.
 From analysis of
Supernovae Type-Ia, Cosmic Chronometers, and Gamma-ray bursts, it
was also found that $\Omega_m\approx 0.87$ for the VFD model
\cite{Haridasu}.\footnote{However, when one compares
$\Omega_b=\frac{8\pi G \,\rho_b}{3H^2 }$ from nucleosynthesis with
$\Omega_m$ from cosmological observations, the result could depend
on the possible renormalization of the gravitational constant
\cite{universe8090456}. Then, the gravitational constant measured
on the Earth or the solar system can differ from the constant used
in cosmology for the uniform universe.} It means that there is no
need for any ad hoc dark matter in the VFD cosmological model
because $\Omega_m\approx\Omega_b$. Moreover, as it was conjectured
in \cite{universe8090456}, spatially nonuniform vacuum
polarization should be taken into account in the dynamics of the
structure formation.

On the other hand, there is a lot of time for the growth of
inhomogeneities in the VFD model \cite{eqofst1,equation}, and the
nonlinear regime begins soon after the last scattering surface.
That allows the suggestion that almost all the baryonic matter
collapses into eicheons \cite{Eicheons}, which replace the black
holes in FVT. There are no strong constraints on the abundance of
black holes in a region of mass $M\sim
10^{13}${--} 
$10^{19}~~M_{\bigodot}$ \cite{Carr2022}, and it is possible that
the matter concentrates namely in this region.

In the VFD model, there is no cosmological deuterium production.
The amount of lithium is less than that in the $\Lambda$CDM, that
alleviates the lithium overproduction problem of the standard
cosmological model.   The amount of $\mbox{CNO}$ is $10^7$ times
greater compared to $\Lambda$CDM, but it does not contradict the
observations \cite{Cassisi,Coc2014}.

\vspace{-6pt}

\begin{table}[b]
\caption{\label{nn1}%
Final abundances of light elements in the $\Lambda$CDM model at
$\Omega_b=0.049$ and the VFD model at $\Omega_b=0.87$. }
\begin{ruledtabular}
\begin{tabular}{ccc}
\textbf{  }& {\boldmath{$\Lambda$}\textbf{CDM}}&
{\textbf{VFD}}\\
\colrule
H     & 0.75         & $0.75$  \\
$\mbox{Y}_p=4\mbox{Y}_{\mbox{He}}$     &   $0.25$        &  $0.25$  \\
$\mbox{D/H}\times10^5$     &     2.6       & $< 10^{-25}$ \\
$^3\mbox{He/H}\times10^5$     &     1.1       & $< 10^{-8}$ \\
$\mbox{T/H}\times10^8$     &    7.9       & $ <10^{-32}$ \\
$(^7\mbox{Li}+^7\mbox{Be})/\mbox{H}\times {10}^{10}$   &  5.7   &   $2.1$   \\
$^6\mbox{Li/H}\times {10}^{14}$ & 1.2& $<10^{-25}$\\
$^9\mbox{Be/H}\times{10}^{19}$ & 9.2 & $<10^{-34}$ \\
 $^{10}\mbox{B/H}\times{10}^{21}$ & 2.9 & $<10^{-8}$\\
 $^{11}\mbox{B/H}\times{10}^{16}$ & 3.3 & $<10^{-10}$ \\
$\mbox{CNO/H}\times {10}^{16}$& 8.0  & $5.6\times 10^7$ \\
\end{tabular}
\end{ruledtabular}
\end{table}

It is widely believed that deuterium is produced only
cosmologically in $\Lambda$CDM, but for the VFD model, the most
plausible and direct way is to create necessary deuterium  by
beams of antineutrino arising during a collapse \cite{Woo} before
the formation of eicheons in the range of $M\sim
10^{13}${--}$10^{19}~~M_{\bigodot}$.
 Their formation is
  unavoidable in the slowly expanding cosmologies because there is much time for the collapse of
  inhomogeneities in contrast to the standard cosmological model.
Indeed, the matter stored in the supermassive eicheons is not
related to ``dark matter'' observed in rotational curves of
galaxies because the latter could be explained by the vacuum
polarization \cite{universe8090456}. Other mechanisms of
non-cosmological deuterium production are also discussed
\cite{Jedamzik2002}.

\subsection{Notes about Cosmic Microwave Background in the Slowly Expanding Cosmological Models}

By this time, there are no trustable studies of the CMB background
for slowly expanding cosmological models, and only some heuristic
calculations exist \cite{Cherkas2018}. The main question is the
origin of a scale corresponding to the first peak in the CMB
spectrum and the origin of the baryon acoustic oscillations (BAO)
ruler  \cite{Fujii2020}. In the standard cosmological model, this
is the sound horizon's size at the recombination moment. For the
Milne-like cosmology, these quantities must be different
\cite{Tutusaus2016,Fujii2020}. Apart from this, the sound horizon
for the Milne-like flat universe is vast and cannot be a scale,
which determines the position of the first CMB peak. Let us
hypothesize that the width of the last scattering surface
\cite{Cherkas2018} could be such a scale for VFD.\footnote{In
$\Lambda$CDM, the recombination turns out to be almost
instantaneous, i.e., the last scattering surface is very thin.} In
this light, the mechanisms of perturbation growth during a
recombination period are of interest \cite{Lewis2007}. As for the
BAO ruler, it has to be determined by the complex nonlinear
process in the slowly evolving cosmologies and is not related
directly to the scale corresponding to the first peak of CMB.

\section{Size of Eicheon}

The concept of \ae ther considered in this work is based on
postulating the preferred coordinate frame, namely, CUM. One more
consequence of this hypothesis is a replacement of the black hole
solutions of GR by the so-called ``eicheons''. In Ref.
\cite{Eicheons}, the spherically symmetric solution of the
Einstein equations in the CUM metrics (\ref{interv1}) was
analyzed, and it was found that the finite pressure solution
exists for an arbitrarily large mass. As a result, there are no
compact objects with an event horizon,\footnote{The event horizon
is a region of space-time that is causality disjointed from the
rest of space-time.} because an ``eicheon'' appears instead of a
black hole \cite{Eicheons}.\footnote{The observations revealed the
phenomena such as ultra-speed star motion, accretion disks around
the super-massive and extremely compact objects (e.g., see
\cite{gillessen2009monitoring,akiyama2021first}), and
gravitational waves from colliding compact objects of stellar mass
\cite{abbott2016properties}, which fit well in the black hole
concept. However, the claims about ``black hole discovery'' should
be treated with caution because these observations do not rule out
completely the alternative theories (e.g., see \cite{K2016350}),
which also admit the existence of extremely compact massive
objects with the exterior mimicking a black hole.}

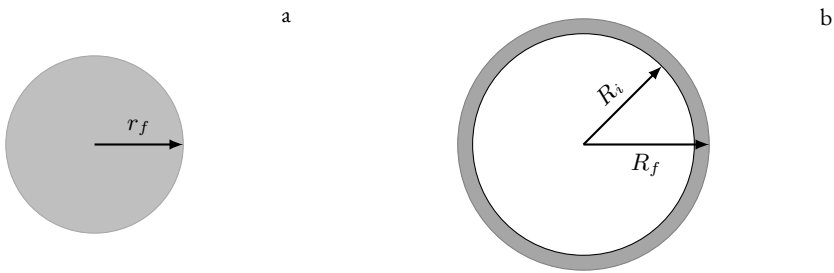
\begin{figure}[H]
\begin{tikzpicture}
\draw[gray!70,fill=gray!50] (0,0) circle (1.7*0.69379cm);

  \draw (1.7*1.5,1.7*1) node {a};
  \draw (1.7*6.9-2,1.7*1) node {b};
 \draw[->,  thick,  arrows={-latex}]  (0,0) -- (1.7*0.69379,0) node[sloped,midway,above] {$r_f$};
 \draw[gray,fill=gray!70] (1.7*5-2,0) circle (1.7*0.98358cm);
\draw[black,fill=white] (1.7*5-2,0) circle (1.7*0.867064cm);
 \draw[->,  thick,  arrows={-latex}]  (1.7*5-2,0) -- (1.7*5.98358-2,0) node[sloped,midway,below] {$R_f$};
 \draw[->,  thick,  arrows={-latex}]  (1.7*5-2,0) -- (1.7*5.61311-2,1.7*0.61311) node[sloped,midway,above] {$R_i$};
\end{tikzpicture}
\caption {\normalfont (\textbf{a}) A  compact object of
uncompressible fluid  with the radius of $r_f$ in the CUM metrics
(\ref {eq42}) looks as a shell (\textbf{b}) with the boundaries
$r_g<R _i<R_f$  in Schwarzschild's type metric, where $r_g$ is a
Schwarzschild's radius.}
\label{fig31}
\end{figure}

In Ref. \cite{Eicheons}, we have turned from the CUM metrics
(\ref{eq42}) to Schwarzschild-like in order to demonstrate that a
compact object looks like a hollow sphere with a radius greater
than that of Schwarzschild (see Figure \ref{fig31}).\vspace{-6pt}

Here, we intend to calculate the radius of a compact object of
constant density in the CUM metrics depending on maximum pressure
and density. For a spherically symmetric space-time, the CUM
metrics (\ref{interv1}) is reduced to\vspace{-6pt}
\begin{equation}
ds^2 = a^2(d\eta ^2 - \tilde {\gamma }_{ij} dx^idx^j) = e^{2\alpha
}\left( {d\eta ^2 - e^{ - 2\lambda }(d{\bm x})^2 - (e^{4\lambda }
- e^{ - 2\lambda })({\bm x}d{\bm x})^2 / r^2} \right),
\label{eq31}
\end{equation}
where $r=|\bm x|$ and $\alpha$, $\lambda$ are the functions of
$r$. In the spherical coordinates, Equation (\ref{eq31}) looks as
\begin{equation}
ds^2 = e^{2\alpha }\left( {d\eta ^2 - dr^2e^{4\lambda } - e^{ -
2\lambda }r^2\left( {d\theta ^2 + \sin^2\theta d\phi ^2} \right)}
\right){\kern 1pt} {\kern 1pt}.
\label{eq42}
\end{equation}

Let us compare (\ref{eq42}) with Schwarzschild's type
metrics\vspace{-6pt}
\begin{equation}
ds^2=B(R)dt^2-A(R)dR^2-R^2\left( {d\theta ^2 + \sin^2\theta d\phi
^2} \right).
\label{ab}
\end{equation}

The difference between the metrics (\ref{eq31}) and (\ref{ab}) is
that the metric (\ref{ab}) suggests that the circumference equals
$2\pi R$. However, there is no evidence for this fact in an
arbitrary spherically symmetric space-time. For the metric
(\ref{eq31}), the circumference is not equal to $2\pi r$ in the
close vicinity of a point-like mass. Coordinate transformation
$t=\eta$, $R=R(r)$ relates the metrics (\ref{eq42}) and
(\ref{ab}), while their comparison gives:
\begin{equation}
B(R)=e^{2\alpha },~~~~~~~~~~~~~~~~~~~\label{e1}
\end{equation}
\begin{equation}
R^2=r^2e^{-2\lambda+2\alpha},~~~~~~~~~~~~~~~~\label{e2}
\end{equation}
\begin{equation}
 A(R)\left(\frac{dR}{dr}\right)^2=e^{4\lambda+2\alpha}\label{e3}.
\end{equation}

Using (\ref{e1}), (\ref{e2}) in (\ref{e3}) to exclude $\lambda$
and $\alpha$ yields
\begin{equation}
\label{eqr}
\frac{dr}{dR}=\frac{R^2}{r^2}\frac{A^{1/2}}{B^{3/2}}.
\end{equation}

For an empty Schwarzschild space-time $A(R)=(1 - r_g/R)^{-1}$ and
$B(R)=1 - r_g/R$, whereas in the region filled by matter, $A(R)$
and $B(R)$ obey \cite{wein}\vspace{-6pt}
\begin{equation}
\label{eqA}
\frac{d}{dR}\left(\frac{R}{A}\right)=1-\frac{6}{M_p^2}\rho R^2,
\end{equation}
\begin{equation}
\label{eqB}
\frac{1}{B}\frac{dB}{dR}=-\frac{2}{p+\rho}\frac{dp}{dR},
\end{equation}
where $r_g=\frac{3 m}{2\pi M_p^2}$. Further, as in
\cite{Eicheons}, we will consider a model of the constant density
$\rho(R)=\rho_0$. In this case, Equations (\ref{eqA}) and
(\ref{eqB}) can be integrated explicitly that gives\vspace{-6pt}
\begin{equation}
A=\frac{R}{R-r_g-2 \rho_0 \left(R^3-R_f^3\right)M_p^{-2}},
\end{equation}
\begin{equation}
B=\left(1-\frac{r_g}{R_f}\right)\frac{\rho_0^2}{(p(R)+\rho_0)^2}
\label{exB}
\end{equation}
 and one needs only
to find a pressure $p(R)$, which obeys the
Tolman--Volkov--Oppenheimer (TVO) equation\vspace{-6pt}
\begin{equation}
p^\prime(R)=-\frac{3}{4\pi M_p^2R^2} \mathcal M(R) \rho (R)
\left(1+\frac{4 \pi R^3 p(R)}{\mathcal M(R)}\right)
\left(1+\frac{p(R)}{\rho (R)}\right)\left(1-\frac{3 \mathcal
M(R)}{2\pi M_p^2 R}\right)^{-1}.
\label{op}
\end{equation}

It is convenient to measure density and pressure in the units of
$M_p^2r_g^{-2}$,  so that  the mean density of Schwarzschild black
hole $\rho_0=m/(\frac{4}{3}\pi r_g^3)$  equals $1/2$, while the
TOV limit $R_f<\frac{9}{8}r_g$ gives the value of
$\rho_0=\frac{1}{2}\left(\frac{8}{9}\right)^3\approx 0.35$. As for
the eicheon radius in Schwarzschild's  type metric, it equals
$R_f=\sqrt[3]{R_i^3+\frac{1}{2\rho_0}}$ in the units of $r_g$,
where $R_i$ is an inner radius, which determines maximum pressure.
Using
\begin{equation}
\mathcal M(R)=\frac{4\pi}{3}\rho_0\left({R}^3-{R_i}^3\right),
\label{mass}
\end{equation}
for solving the TOV equation for pressure, it is possible to find
$B$, and then solve (\ref{eqr}) with the initial condition
$r(R_i)=0$ and find the eicheon radius  $r_f=r(R_f)$  in the CUM
metrics.

\begin{figure}[H]
\hspace{1 cm}\includegraphics[width=8cm]{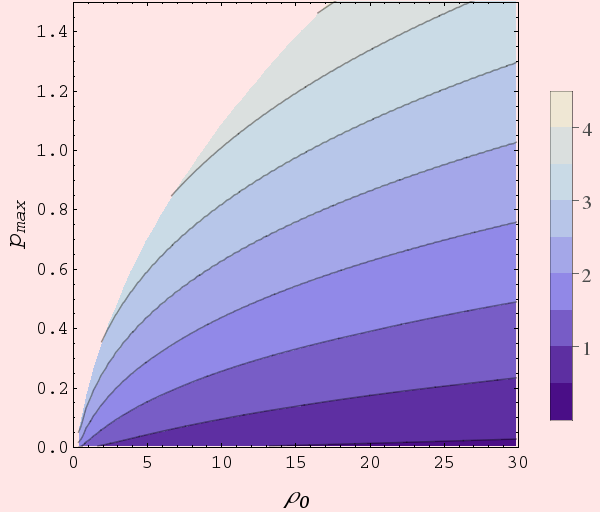} \vspace{0.0 cm}
\caption{Dependence of the eicheon radius $r_f=r(R_f)$ in the CUM
metrics, expressed in the units of gravitational radius, on the
density and maximal pressure (i.e., pressure in the eicheon
center). Pressure and density are in the units of $M_p^2r_g^{-2}$.
}
\label{rad}
\end{figure}

 Let us plot (see {Figure} \ref{rad}) the calculated radius of the
eicheon in the CUM metrics in dependence on density $\rho_0$ and
maximum pressure, that is, the pressure in the center of a solid
ball in the metric (\ref{eq42}). An approaching $R_i$ to unity
increases the maximal pressure. Actual density and pressure in the
center of eicheon are defined by the extremal equation of state,
which is the subject of future investigations. However, Figure
\ref{rad} allows concluding that the pressure is considerably
smaller than the energy density in a region of interest. That
results in a straightforward analytic estimation of the eicheon
radius. For the estimation, one could take pressure equal to some
constant (e.g., $p(R)=\rho_0/10$) in (\ref{exB}), or even simply
$p(R)=0$. Then one could take $R_i=1$, i.e., the Schwarzschild
radius and integrate (\ref{eqr}) \linebreak to obtain
\begin{equation}
r_f=\sqrt[3]{3\int_1^{R_f}\frac{A^{1/2}}{B^{3/2}}R^2dR} \approx
\frac{\sqrt{3} \sqrt[3]{11} \,\rho_0^{1/6}}{2^{5/6}},
\label{rff0}
\end{equation}
where a small ``thickness'' of the eicheon  surface $R_f-1$ is
used, $R_f$ is expressed as  \linebreak
$R_f=\sqrt[3]{1+\frac{1}{2\rho_0}}$ , and only asymptotic term of
large $\rho_0$ is retained. In the ordinary units, the result
reads
\begin{equation}
r_f=\frac{3\ 3^{5/6} \sqrt[3]{11}\, m^{4/3} \sqrt[6]{\rho_0}}{4
\sqrt[6]{2}\, \pi ^{4/3} M_p^3}
\label{rff}
\end{equation}
and it is slightly unexpected because the eicheon radius rises
with the density that turns out to be a specific manifestation of
the CUM geometry.\footnote{Here, we obtain primitive geometrical
formulas connecting the radius of a compact astrophysical object
with its mass and density. To obtain nontrivial formulas
expressing the radius of the object through its mass only, using
the physical equation of state is needed, e.g., nucleonic matter
or strange quark matter as it was done in the neutron star physics
\cite{Lattimer_2001}.} In particular, the eicheon of the Planck
density $\rho_0=M_p^4$, which is sometimes considered as a maximal
density in \linebreak nature \cite{barrau2014planck} has a radius
of $r_f\approx 0.8\frac{1}{M_p}\left(\frac{m}{M_p}\right)^{4/3}$
in the CUM metrics. Looking at the last equation, one may assume
that the large eicheons cannot be very dense. However, $r_f$ given
by (\ref{rff0}) is not a physical distance but only points out a
border of eicheon in the CUM metrics, whereas the physical
distance is given by $l_{eiche}=\int_0^{r_f}e^{\alpha+2\lambda}dr
=\int_1^{R_f}A^{1/2}dR \approx \frac{5}{24\rho_0}$.

Recently, many investigations explored the footprints of black
holes manifesting themselves through star trajectories and a
shadow in the accretion disks around the galaxy centers,
gravitational lensing, and gravitational waves from the colliding
compact objects (see footnote 3 on p. {15}
). These phenomena can be explained from the properties of both
stationary and non-stationary metrics of Schwarzschild and Kerr
types, where the radius of ``source'' objects is of the order of
$r_g$. It seems reasonable to interpret these observations in the
CUM framework and  obtain an actual eicheon radius $r_f$ using its
equation of state.\footnote{It could be compared with properties
of neutron and exotic stars \cite{sandin2007compact}.} The most
informative study originates  from collisions of ultracompact
massive objects producing the gravitational waves observed by the
existing and developing detectors. At this moment, direct
astrophysical observations and, much less, analogous modeling
cannot provide decisive evidence, which would rule out some
alternative concepts of ultracompact massive objects without a
horizon (nevertheless, see \cite{Bambi2017}). One could suggest
that off-horizon properties of an eicheon are the same as for a
black hole. However, near-horizon phenomena like gravitation wave
emission in black hole collisions could be most informative
\cite{K2016350,PhysRevLett041302,PhysRevD024058} and the study of
these phenomena in the framework of FVT  is a matter for the
future.

\section{Decoherence of the Particles Due to
Gravitational Potential Fluctuations}

Here, we return to the consideration of a locally defined scale
factor as an operator. One more implication of the CUM metrics and
\ae ther arises for a gravitational \linebreak decoherence
\cite{cherkas2021wave}, which is the subject of the table-top
quantum gravity experiments.
 In GR, it would not
be possible to say that the vacuum fluctuations under Minkowski's
space-time are small. Actually, for the small vacuum fluctuations,
one could turn to the reference system where they are significant.
That means the appearance of the so-called gauge waves, which are
the consequence of the reference frame choice. By restricting the
possible reference systems, it would be possible to reveal the
actual vacuum fluctuation influencing the motion of the massive
particles. The main fundamental question is: Does a massive
particle lose its coherence due to interaction with \ae ther?
Under Minkowski's background, one could write for a locally
defined scale factor:\vspace{-6pt}
\begin{equation}
a(\eta,\bm r)=(1+\Phi(\eta,\bm r)),
\end{equation}

According to \cite{cherkas2021wave}, the correlator of the Fourier
amplitudes for the gravitational potential in vacuum
 $\hat \Phi(\eta,\bm r)=\sum_{\bm k} \hat \Phi_{\bm k}(\eta)e^{i\bm k\bm r}$ takes the form
\begin{equation}
S(\tau-\eta,k)=<0|\hat \Phi_{\bm k}^+(\eta)\hat \Phi_{\bm
k}(\tau)|0>=\int_{-\infty}^{\infty} \tilde
S(\omega,k)e^{i\omega(\tau-\eta)} d\omega,
   \label{sigm2}
   \end{equation}
where  a spectral function $\tilde S(\omega,\bm q)$  is
approximately written as \cite{cherkas2021wave}
\begin{equation}
 \tilde S(\omega,k)\approx \left\{\begin{array}{c}
 \frac{N_{all}}{32  \pi ^2 {M_p}^4}, \,q<\omega<2k_{max} \\
 0,  \,\,\mbox{otherwise}.
\end{array}\right.,
\label{spectrum}
\end{equation}
where $N_{all}=N_{boson}+N_{ferm}$ is a number of all degrees of
freedom.

For a nonrelativistic massive point particle propagating among the
fluctuations of the gravitational potential, the Fokker--Plank
equation is:
\begin{equation}
\ptl_\eta f_{\bm k}(\bm p)+i(E_{{\bm p}+{\bm k}/2}-E_{{\bm p}-{\bm
k}/2})f_{\bm k}(\bm p)= -i\,K_1\,\bm k\frac{\ptl f_{\bm k}}{\ptl
\bm p} +2i K_{2}\,\bm k\bm p\,\Delta_{\bm p}f_{\bm k}(\bm p) +2 i
K_{3}\,p_i k_j \frac{\ptl^2 f_{\bm k}}{\ptl p_j\ptl p_i},~~
\label{fokker}
\end{equation}
where $ f_{\bm k}(\bm p)$ is a Fourier transform of the Wigner
function $\tilde f(\bm r,\bm p)=\sum_{\bm k}f_{\bm k}(\bm
p)e^{i\bm k\bm r}$. In the first order on the constants $K_1$,
$K_2$, $K_3$ it is possible to write:\vspace{-6pt}
\begin{equation}
\int  f_{\bm k}(\bm p,\eta)f_{-\bm k}(\bm p,\eta)d^3\bm pd^3\bm
k\approx 1-(3 {K_1}+3 {K_2}+6 {K_3})\frac{\Gamma ^2 \eta^2 }{m}.
\label{resdec}
\end{equation}

  {This means that the interaction with a vacuum
produces decoherence manifesting itself in the decreasing of
``purity'' \cite{cherkas2021wave} of a particle state according to
(\ref{resdec}). From Equation~(\ref{resdec}), the decoherence time
is estimated as}
\begin{equation}
t_{dec}\approx\frac{1}{\Gamma}\sqrt{\frac{m}{3 {K_1}+3 {K_2}+6
{K_3}}},
\label{tdec}
\end{equation}
and using the approximate expressions for the constants $K_1, K_2,
K_3$, the decoherence length can be found \cite{cherkas2021wave}
\begin{equation}
L_{dec}\approx \frac{4 M_p}{3\sqrt{3
\,N_{all}}\pi\,m}\frac{v}{\Gamma}\,,
\label{ldec}
\end{equation}
where $v$ is a particle velocity, $m$ is a particle mass,
$1/\Gamma$ is a localization length of the particle wave packet.
That is, a point-like particle of mass $ m\sim \frac{4 M_p\,
v}{3\sqrt{3 N_{all}}\pi}$ loses coherence at a distance equal to
the length of the wave packet $1/\Gamma$. It should be noted that
interaction with the \ae ther does not produce a particle
scattering because the momentum distribution $f_0(\bm p)$ does not
change. Nevertheless, decoherence arises. That is a fundamental
result implying Lorentz and Galilean invariance violation because
one particle state becomes non-pure quantum state.

The difference in particle propagation in the QG and QFT is
illustrated in Figure~\ref{vv}. The \ae ther in QG originates from
an absence of an invariant vacuum state. The last is not invariant
relatively to the general transformation of coordinates and, in
particular relative to the Lorentz transformation when it is
considered as a subgroup of the general transformation of
coordinates.

As regards the decoherence observation (\ref{ldec}), such massive
point particles are unknown. A real particle of large mass has a
finite size, which restricts momentums transferred by the particle
form factor: $q<1/d$, where $d$ is particle size. In this case,
the following estimation arises
\cite{cherkas2021wave}\vspace{-6pt}
\begin{equation} L_{dec}\approx\frac{8\pi (M_p
d)^2}{\sqrt{N_{all}}} \frac{v}{\Gamma}.\end{equation}
 This quantity seems very large
and unobservable. At the same time, the real particles are not
rigid but have internal degrees of freedom and consist of a number
of point particles, so more careful investigation is needed.
Moreover, other possible fundamental mechanisms of decoherence
also need investigation \cite{dec0}.

Recently, a gravitationally induced entanglement has attracted
great attention (see e.g.,
\cite{bel,carney2022newton,dan,rov,krisnanda2020observable}).
There is no doubt that the nonrelativistic quantum mechanics holds
for any interaction, including gravitational interaction in the
form of the second Newton's law and any weak external
gravitational field \cite{gorb}. It is also no doubt that the
gravitational waves of linearized gravity are fully analogous to
electromagnetic waves and have to be quantized. Undoubtedly, the
second Newton's law could be interpreted as an exchange by
gravitons, like the Coulomb law can be interpreted as an exchange
by photons. In contrast, the \linebreak result (\ref{ldec}) seems
much less trivial because this fundamental decoherence implies the
existence of \ae ther with stochastic properties. Such an \ae ther
is absent in quantum electrodynamics due to the existence of the
LI vacuum state. Implications of the  \ae ther in a photon sector
of LI violation \cite{sym12081232,arxiv2111} have to be
investigated.

\section{Conclusions}

To summarize, the CUM metrics gives a sustained basis for quantum
gravity physics, cosmology, and physics of compact astrophysical
objects. Although fascinating physics like closed time-like curves
\cite{closed1,5801002B,Ahmed}, time machines
\cite{nov,wuthrich2019time}, wormholes \cite{Teo1998}, and Hawking
\linebreak radiation
\cite{herrero2021hawking,Coogan2021,saraswat2021extracting} are
excluded in the CUM metrics, these metrics give a fresh impetus to
investigate the real physical phenomena, including the structure
formation \cite{equation}, CMB \cite{Cherkas2018}, the structure
of ultracompact astrophysical objects, and search for the
decoherence QG effects and other QG consequences from the vacuum
fluctuations of the gravitational potential. All these phenomena
 imply to single out the conformally unimodular metrics corresponding
  to a reference system, where the \ae ther is at rest ``in tote''. Certainly,
  it suggests the \ae ther existence per se.
In QG, an \ae ther is not simply some background but a thing that
weaves all the physical phenomena into a whole quantum universe.

On the other hand, because black holes are absent in this theory,
there is no actual ``eraser'' of the information in the CUM
metrics. In other words, a wave function of some particular
quantum system is only mixed to a more general wave function,
including vacuum and, finally, all universe, and  the universe's
wave function seems not an idealization, but a reality conserving
all information without any loss \cite{inform}. \vspace{6pt}

\appendix
\section{Quasi-Heisenberg Quantization}
\label{quasi}

For simplicity, it is convenient to consider two scalar fields,
$\phi_1$ and $\phi_2$, that correspond to a system with three
degrees of freedom, including the logarithm of scale factor
$\alpha=\ln a$. As a result, there is only one degree of freedom
because the Hamiltonian and momentum constraints allow excluding
two of them. Let us discuss a quantum picture of the \linebreak
system
  (\ref{mod}), (\ref{sv0}), and (\ref{sv}). The quasi-Heisenberg picture suggests that one needs to define the commutation relations and initial values for operators at the initial moment and then permit the operator evolution according to the equation of motions.
For quantization with the help of the Dirac brackets (see also
\cite{concept}), one should set two additional gauge fixing
conditions corresponding to the Hamiltonian and momentum
constraints.

  Let us take these conditions as
  \begin{eqnarray}
\hat \alpha(0,\sigma)=\alpha_0=const, \label{condal}\\
\ptl_\sigma \hat {\pi}_1(0,\sigma)=0,\label{condB}
\end{eqnarray}
i.e., the logarithm of the scale factor and momentum $\hat
{\pi}_1(0,\sigma)=\pi_1$ are $c$-number constants at the initial
moment. Generally, that is some time-dependent gauge, which is
known only at an initial moment. Then it is permissible for the
commutation relations to evolve.

Dirac brackets could allow calculating the operator commutation
relations  at the initial moment, but the equivalent receipt is to
set\vspace{-6pt}
\begin{equation}
\hat
\phi_2(0,\sigma)\equiv\varphi(\sigma),\label{cond2}\end{equation}
\begin{equation}
\hat \pi_2(0,\sigma)\equiv-i\frac{\delta}{\delta
\varphi(\sigma)},~~~~~\hat \phi_2^\prime(0,\sigma)=-i
\,e^{-2\alpha_0}\frac{\delta }{\delta
\varphi(\sigma)}\label{cond3}
\end{equation}
and express other variables from constraints and gauge conditions
to obtain
\begin{equation}
\hat p_\alpha(0,\sigma)=\sqrt{-\frac{\delta^2}{\delta
\varphi^2(\sigma)}+\pi_1^2},~~~~\hat
\alpha^\prime(0,x)=e^{-2\alpha_0}\sqrt{-\frac{\delta^2 }{\delta
\varphi^2(\sigma)}+\pi_1^2}, \label{rlz2}
\end{equation}
\begin{equation}
\hat \phi_1(0,\sigma)=\frac{i}{\pi_1}\int_{-\infty}^\infty
\theta(\sigma-\sigma^\prime)S\left( \frac{\delta}{\delta
\varphi(\sigma^\prime)}\,\ptl_{\sigma^\prime}\varphi(\sigma^\prime)\right)d\sigma^\prime,\label{cond4}
\end{equation}
\begin{equation}
\hat \phi_1^\prime(0,\sigma)=e^{-2\alpha_0}\pi_1,\label{cond5}
\end{equation}
 where the symbol $S$ denotes symmetrization of the
noncommutative operators, i.e., \linebreak $S(\hat A \hat
B)=\frac{1}{2}(\hat A\hat B+\hat B\hat A )$ or $S(\hat A\hat B\hat
C)=\frac{1}{6}(\hat A\hat B\hat C+\hat B\hat A\hat C+\hat A\hat
C\hat B+\dots)$ and $\theta(\sigma)$ is a unit step function.  Its
appearance in (\ref{cond4}) is the only nontrivial moment that
follows from calculation of the Dirac brackets \cite{bia}, and we
have introduced it here for expressing $\phi_1$ from the momentum
constraint (\ref{sv}).

The equations of motion (\ref{eqns10}), (\ref{eqns1})  should be
considered as the operator equations with the initial conditions
(\ref{condal}), (\ref{cond2}){--} (\ref{cond5}). The second stage
of quantization consists of building the Hilbert space where the
quasi-Heisenberg operators act. This stage again begins from the
classical Hamiltonian (\ref{sv0})  and momentum (\ref{sv})
constraints. The momentum constraint and corresponding gauge
condition (\ref{condB}) are resolved relatively the variable
$\phi_1$ and its momentum $\pi_1$. Then, these quantities are
substituted to the Hamiltonian constraint, which is then quantized
and considered as the Wheeler--DeWitt equation in the vicinity of
the small scale factor \emph{a}$\sim$0, i.e., $\ln
a=\alpha\rightarrow-\infty$. In such a way, we come to
\begin{equation}
\left(\frac{\delta^2}{\delta
\alpha(\sigma)}-\frac{\delta^2}{\delta^2
\varphi(\sigma)}+\pi_1^2\right)\Psi[\alpha,V]=0, \label{witt}
\end{equation}
where it is taken into account that $\pi_1$ is some constant.
Space of the negative frequency solutions of the equation
(\ref{witt}) constitutes the Hilbert space for the
quasi-Heisenberg operators.

In the general case, the solution of Equation (\ref{witt}) is of
the form of the wave packet
\begin{equation}
\Psi[\alpha,\varphi]=\int C[k]\,e^{\int\left(-i
\alpha(\sigma)\sqrt{\pi_1^2+k^2(\sigma) }+ik(\sigma)
\varphi(\sigma)\right)d \sigma}\,\mathcal D k(\sigma),
\end{equation}
where only negative frequency solutions are taken and  $\mathcal D
k(\sigma)$ denotes a functional integration over $k(\sigma)$. The
scalar product has a form
 \cite{prep1,cherkas2005quantum,toy}
\begin{eqnarray}
   <\Psi|\Psi>=i\, Z \prod_\sigma\int \biggl( \Psi^*[\alpha,\varphi] {\hat
D^{-1/2}}(\sigma)\frac{\delta}{\delta \alpha(\sigma)
}\Psi[\alpha,\varphi]\nonumber\\~~~~~~~~~~~~~~~~~~-\left({\hat
D^{-1/2}}(\sigma)\frac{\delta}{\delta \alpha(\sigma)
}\Psi^*[\alpha,\varphi]\right)\Psi[\alpha,\varphi]\biggr)d
\varphi(\sigma),
\label{scal}
\end{eqnarray}
where $\hat D(\sigma)=-\frac{\delta^2}{\delta
\varphi^2(\sigma)}+\pi_1^2$ and $Z$ is a normalization constant.
The infinite product is taken over $\sigma$-points, and to be
understood in a formal sense as representing the result of a
limiting process based on a lattice in $\sigma$-space. The scalar
product (\ref{scal}) is independent of the choice of the
hyperplane $\alpha(\sigma)$.

The mean value of an arbitrary operator can be evaluated
as\vspace{-6pt}
\begin{eqnarray}
   <\Psi|\hat A[\alpha,-i\frac{\delta}{\delta \varphi(\sigma)},\varphi(\sigma)]|\Psi>=i \,Z
\prod_\sigma\int \biggl( \Psi^*[\alpha,\varphi] \hat A\,{\hat
D^{-1/2}}(\sigma)\frac{\delta}{\delta \alpha(\sigma)
}\Psi[\alpha,\varphi]\nonumber\\~~~~~~~~~~~~~~~~~~-\left({\hat
D^{-1/2}}(\sigma)\frac{\delta}{\delta \alpha(\sigma)
}\Psi^*[\alpha,\varphi]\right)\hat A\,\Psi[\alpha,\varphi]\biggr)d
\varphi(\sigma)\biggr|_{\,\alpha(\sigma)=\alpha_0\rightarrow
-\infty }.
\label{mean1}
\end{eqnarray}

Let us note that the hyperplane $\alpha(\sigma)=\alpha_0$ along
which the integration is performed \linebreak in (\ref{mean1}), is
the same as it is used as an initial condition for the
quasi-Heisenberg operator $\hat \alpha$ in (17). In a more
convenient momentum representation $\hat \pi_2(\sigma)=k(\sigma)$,
$\hat \phi_2(\sigma)=i\frac{\delta }{\delta k(\sigma)} $, the wave
function $\psi$ is\vspace{-6pt}
\begin{equation}
{\psi}[\alpha,k]=C[k]\exp\left({-i\int\alpha(\sigma)\sqrt{k^2(x)+\pi_1^2}\,dx}\right).
\end{equation}
Then, the mean value of an operator becomes\vspace{-6pt}
\begin{eqnarray}
  <\psi|\hat A[\alpha(\sigma),k(\sigma),i\frac{\delta}{\delta
k(\sigma)}|\psi>=~~~
 ~~~~~~~~~~~~~~~~~~~
 ~~~~~~~~~~~~~~~~~~~~~~~~~~~~~~~~\nonumber\\\int
C^*[k]e^{-i \int \alpha(\sigma)\sqrt{k^2(\sigma)+\pi_1^2}
\,d\sigma}\hat A\,e^{i \int
\alpha(\sigma)\sqrt{k^2(\sigma)+\pi_1^2}\,d\sigma}C[k]\,\mathcal D
k(\sigma)\biggl|_{\,\alpha(\sigma)=\alpha_0\rightarrow -\infty }.
\label{mean2}
\end{eqnarray}

Thus, we have an exact quantization scheme consisting of the
Wheeler--DeWitt equation in the vicinity of small scale-factor
(\ref{witt}), the operator initial conditions (\ref{rlz2}) for the
equations of motion and the expressions (\ref{mean1}) {and}
(\ref{mean2}) for calculation of the mean values of operators.

\bibliography{particles4}

\end{document}